# A consequence of failed sequential learning: A computational account of developmental amnesia

Qi Zhang

Madison, WI, USA
qzhangsensor@gmail.com

**Abstract**

Developmental amnesia, featured with severely impaired episodic memory and almost normal semantic memory, has been discovered to occur in children with hippocampal atrophy. This unique combination of characteristics seems to challenge the understanding that early loss of episodic memory may impede cognitive development and result in severe mental retardation. Although a few underlying mechanisms have been suggested, no computational model has been reported that is able to mimic the unique combination of characteristics. In this study, a cognitive system is presented, and developmental amnesia is demonstrated computationally in terms of impaired episodic recall, spared recognition and spared semantic learning. Impaired sequential/spatial learning ability of the hippocampus is suggested to be the cause of such amnesia. Simulation shows that impaired sequential leaning may only result in severe impairment of episodic recall, but affect neither recognition ability nor semantic learning. The spared semantic learning is inline with the view that semantic learning is largely associated with the consolidation of episodic memory, a process in which episodic memory may be mostly activated randomly, instead of sequentially. Furthermore, retrograded amnesia is also simulated, and the result and its mechanism are in agreement with most computational models of amnesia reported previously.

# A consequence of failed sequential learning: A computational account of developmental amnesia

## 1. Introduction

Declarative memory, or explicit memory, can be fractionated into episodic memory and semantic memory (Tulving, 1972). Episodic memory refers to personally based memories and semantic memory refers to the memory of factual knowledge. The loss of the capacity in retaining episodic memory leads to various amnesias, e.g., retrograde amnesia and anterograde amnesia. Neuropsychological analyses and functional imaging studies (e.g., Aggleton & Brown, 1999; Eichenbaum, 1992, Hodges & Carpenter, 1991; Zola et al. 1986) have broadly confirmed the link between declarative memory and the medial temporal lobe (MTL). The MTL consists of the hippocampus (including the CA fields, dentate gyrus and subiculum), and the adjacent entorhinal, perirhinal and parahippocampal cortices, and plays a critical role in both episodic memory retention and semantic memory acquisition through memory consolidation (see Eichenbaum, 2004). Many theoretical and computational models have been proposed to characterize the important role of the MTL in semantic learning via memory consolidation of episodic memory (e.g., Gluck & Myers, 1993; Hasselmo et al., 1996; Milner, 1989; Nadel et al., 2000; Squire et al., 1984; Treves & Rolls, 1992). Some of the models have simulated how a damaged hippocampus may affect the capacity of episodic memory, resulting in retrograde amnesia and/or anterograde amnesia, as well as, the failure of memory consolidation (e.g., Alvarez & Squire, 1994; McClelland *et al*., 1995; Meeter & Murre, 2005; Murre, 1996; O'Reilly & Rudy, 2000). Therefore, lesioned hippocampus may severely impair the acquisition of semantic memory since impaired episodic memory could result in the deficit of memory consolidation. This view is supported by empirical findings about patients with hippocampal amnesia who often show extreme difficulty in the acquisition of semantic knowledge (e.g., Gabrieli et al., 1988; Postle & Corkin, 1998).

The finding of developmental amnesia first reported by Vargha-Khadem et. al. (1997) provides a new twist to the understanding about the correlation between episodic memory and semantic memory. Developmental amnesia is a rather atypical form of memory deficit that has been discovered to occur in children with hippocampal lesion. The patients studied by Vargha-Khadem et. al. had all suffered bilateral damage limited to the hippocampal formation at very early ages with sparing of surrounding cortical areas. Crucially, they all appeared to have profound lack of memory for the events of everyday life, but all developed normal levels of performance on tests of intelligence, language, and general knowledge. Subsequent studies of other developmental amnesic patients (e.g., Baddeley et al., 2001; Isaacs et al. 2003) confirmed the severely impaired episodic memory that is accompanied by intact capacity of semantic learning. The impairment of episodic memory in such patients is unique, and appears to only affect the capacity of recall, not of recognition. For example, Baddeley et al. (2001) reported a patient who appeared to lack the recollective phenomenological experience normally associated with episodic memory, but performed within the normal range on each of six recognition tests.

Early loss of episodic memory should impede cognitive development and result in severe mental retardation, based on the view that semantic memory largely results from the consolidation of episodic memory. This does not seem to be the case for patients with developmental amnesia. Vargha-Khadem et al. (1997) suggested that the unique characteristics of developmental amnesia could be explained based on the view that episodic memory is partially dissociated from semantic memory, and only episodic memory is fully dependent on the hippocampus. They also argued that recognition could be dissociated from recall memory, which was supported by evidence from hippocampectomized monkeys. Baddeley et al. (2001) further suggested that the recollective process of episodic memory is not necessary for either recognition or acquisition of semantic knowledge. On the other hand, Squire & Zola (1998) and Manns & Squire (1999) proposed a number of different explanations. One is that the early onset of damage in the patients may have led to some form of adaptation of semantic acquisition, and another is about the residual episodic memory observed in all such patients. They noted that the damage to episodic memory of the studied patients was not complete (i.e. patients had non-zero scores on recall tests and non-total atrophy in the hippocampus), and suggested this residual ability might be enough to explain their near normal semantic memory performance. While none of the proposed explanations has been commonly agreed on, no computational model has been reported to simulate the unique characteristics of impaired recall, spared recognition and semantic memory under any given underlying mechanism.

In this study, a cognitive and computational model is presented in which the impaired sequential learning of the hippocampus is proposed to be responsible for developmental amnesia. When impaired sequential learning is resulted from the lost association among elements in episodic memory of past experience, the recall of the past experience is impaired, even if all elements of past experience are still stored into episodic memory. However, episodic memory may still be consolidated because sequential activation of episodic memory may not be necessary in the cognitive process. The unique combination of characteristics of developmental amnesia is simulated, and the relationship between memory consolidation rate and time (the opposite of retrograded amnesia) is also simulated.

## 2. The cognitive and computational system

### 2.1. A revisit of a learning system

Semantic memory refers to generic knowledge of the world (Tulving, 1972) that represents knowledge on verbal symbols, their meanings and referents. The flexibility in managing the acquired knowledge is a key characteristic of semantic memory, which has been lacking in many reported computational models. It is largely because these models are not implemented with a practical mechanism to abstract and acquire meanings. The inability to ground symbol (word) with its meaning has been a fundamental obstacle in artificial intelligence system (Harnad, 1990; Smolensky, 1997). In criticizing the paradigm of connectionist modeling, Forster (1994) has argued that because the mapping from word form to meaning is largely arbitrary, and because connectionist networks are not particularly adept at learning arbitrary mappings, they are inappropriate for simulating lexical conceptual processing. Since 1990, researchers have tried different approaches to realize symbol grounding. Among these efforts, a paradigm of symbol grounding has been proposed and implemented in a learning system that can acquire the conceptual knowledge of "zero", "one" and "tally" (Zhang, 2005).

Figure 1

This learning system is adopted to be the semantic system in this study, which is shown in Figure 1 of the combination of the "symbol subsystem" and "representation subsystem". The symbol subsystem learns symbols from external symbol input, the representation subsystem learns the associated common features from external representation input, and they communicate with each other via the "bundle of internal signals". This construct is inspired by the finding of split-brain (Myers & Sperry, 1953) that indicates each brain half appears "to have its own, largely separate, cognitive domain", and to "have its own learning processes and its own separate chain of memories" as described by Sperry in his Nobel lecture (1982). Sperry further noted that our left hemisphere is capable of comprehending printed and spoken word, and our right hemisphere is word-deaf and word-blind, but capable of comprehending spatial and imagistic information.

Table 1

In semantic system of Figure 1, each of the subsystem is a cognitive unit of multi-level information storages and processes. The lowest cognitive level is called "single memory", which has three inputs and four outputs as indicated in Table 1 of "function states". A single memory is a storage unit that permanently stores a signal input ($Isig$) and an excitation input ($Iexc$). It is also a comparator that fires its stored information accordingly after comparing an arriving signal with what has been stored. The comparator function is straight forward, but the learning function needs some more explanation. It takes two steps for a single memory to store (i.e., to learn) an $Isig$/$Iexc$ pair. In the first step, when condition is right as indicated in Table 1, a single memory fires a coordination signal ($Ocor$), and the signal is transported to the "bundle of internal signals". Only when the "bundle" receives a coordination signal from both the symbol and representation subsystems, it generates a unique excitation signal (Iexc) and sends it back to all single memories in both subsystems. In the second step, after the acting single memory receives the Iexc, it stores both the Isig and Iexc. In Figure 1, single memory is shown as a small circle that is marked with 1, 2 or 3.

The next cognitive level above the single memory is called "memory triangle", e.g., the *MTs1* and *MTr1* in Figure 1, which consists of three single memories. In a triangle, the *Oint* of one single memory is the *Iint* of next single memory. As a result, all three single memories form a loop by the interlock signal, *Iint*. The purpose of a triangle is to learn a unique data point (Isig) three times as one single memory learns from input one at a time, based on an understanding that if a data point represents a common feature of external stimuli, it should commonly appear in all external stimuli. After a memory triangle has stored the "common data point" for three times into each of its single memories in the order from 1 through 3, the common feature is considered learned and generalized because of the existence of the loop.

A subsystem with several interlocked memory-triangles is constructed to learn more common features. Since a memory triangle is the base for a common feature, knowledge is locally stored. The localized characteristic has an advantage to self-assemble a logically interrelated structure among acquired knowledge. For example, we have to know the meaning of "zero" before knowing the meaning of "one"; we have to understand what "one" is before knowing what "many" is. These three concepts are interrelated in a hierarchal structure that is simply implemented by those interlock signals as shown in Figure 1. Thus, *MTr1* must learn first, then *MTr2*, and finally *MTr3*, if the to-be-learned common features are logically interrelated.

A concept is an abstract idea or a common feature, and a word is a symbol for concept. A conceptual knowledge is considered fully learned only when a symbol is associated to a common feature. This association is realized by pairing the two subsystems into a semantic system. One subsystem is dedicated for symbolic learning and the other for common feature learning. The association is done by assigning a unique excitation signal (*Iexc*) and storing the *Iexc* together with the symbol and its common feature, which has been explained in the introduction of the single memory. Thus, an excitation signal fired from the symbol subsystem can activate its associated common feature in the representation subsystem, and vice versa.

The outline of either the symbol or representation subsystem, in Figure 1, is called "interface", which is the top cognitive layer of its subsystem and links all its memory triangles. An interface is the information gateway of a subsystem, which delivers external stimuli to its enclosed memory triangles, exchanges information between the two subsystems, and projects signal output to the external world and other subsystems. When an interface receives an external stimulus, it disassembles the stimulus into a sequence of data points, and distributes the sequenced data points to its single memories. It also collects and organizes activated information and forwards them to the opposite subsystem or external world. In this study, the combination of the two subsystems is proposed to resemble the semantic learning function of the neocortex.

Memory formation in a biological system is believed to associate with the changes in synaptic efficiency that permit strengthening of associations between neurons, and activity-dependent synaptic plasticity at appropriate synapses during memory formation is believed to be both necessary and sufficient for storage of information (for a review, see Lynch, 2004). The synaptic efficiency is related to two phases of synaptic modifications: a protein synthesis-independent phase that lasts for hours, and a synthesis-dependent phase that lasts from days to months. In laboratory condition, the short-lived modification can be induced by tetanic stimulation (e.g., Bliss & Lomo, 1973), but the long-lasting modification is mostly induced by a series of tetanic stimulations over a long period of time. The long-lasting modification involves gene transcription (e.g., Impey, et al., 1998) and protein synthesis (e.g., Segal & Murphy, 1998) that enhance existing synapses and form new connections, making it an attractive candidate for the molecular analog of long-term memory (see Lynch, 2004).

The formation of the long-lasting synaptic modification has two critical implications. One is that such formation often requires a good number of continuous stimulations, and the other is that new neural connections may be permanently established after a long-lasting modification. In order to cooperate with the gradual and slow process of long-lasting synaptic modification, an important modification is made to the learning mechanism of the single memory at Step 2 of the learning phase as shown in Table 1. At step 2, a single memory can fire "yes" output of the interlock signal only after it has been stimulated by the same signal input of "Io" for a given number of times over a period of time. As a result, a single memory can only connect to its next single memory in a hierarchal structure of knowledge after it is able to fire "yes" interlock signal. Thus, a "modification delay", $T$, is introduced to every single memory to reflect the "synaptic modification".

## 2.2. The episodic storage

Empirical evidences indicate that the hippocampus and its surrounding areas are indispensable for episodic learning (e.g., Scoville & Milner, 1957). Semantic learning becomes almost impossible when the hippocampus is severely damaged (e.g., Corkin S. 2002). With an additional storage subsystem for episodic learning, the learning system has been demonstrated to be able to learn semantic knowledge from randomly activated past experiences (episodic memory, Zhang, 2009). That study was intended to simulate the functions of learning and memory consolidation in dreams, which have been proposed by numerous functional and neuropsychological studies of dream and dream sleep, while the randomness is the characteristic of dreaming that has long been revealed by Hobson & McCarley initially (1977).

Empirical evidences indicate that developmental amnesia is associated with hippocampal atrophy. Thus, the focus of this study is to simulate some functions of the hippocampus, and to search for the correlation between the unique combination of the three characteristics and the impairment of related functions. The interested functions in this study are sequential learning (e.g., Levy, 1996; Granger et al., 1996; Wallenstein et al., 1998) and spatial navigation (e.g., Burgess et al., 1994; Levy, 1989; McNaughton & Morris, 1987; Sharp, 1991). The Episodic Storage in Figure 1 is designated to implement these two functions, which is developed from the storage subsystem of episodic learning in the previous study for dream functions (Zhang, 2009).

The Episodic Storage consists of a number of memory cells that are enclosed by an interface, which delivers inputs in parallel to all memory cells and collect outputs from them. Similar to a single memory, a memory cell also serves as a storage and comparator. Each memory cell has four inputs (symbol input of *Is-sig*, representation input of *Ir-sig*, interlock input of *Iint*, and firing trigger of *Ifire*) and four outputs (symbol output of *Os-sig*, representation output of *Or-sig*, interlock output of *Oint*, and comparison output of *Ocom*), as shown for "*H2*" in Figure 1. The function states of a memory cell are given in Table 2, which indicates how a memory cell learns, fires and compares. Since all memory cells are interlocked by interlock signals in one direction, a sequence of external events can be both stored and retrieved in their original order of arrival. When a memory cell receives a signal that matches any one of the two stored signals, it fires an *Ocom* of "yes", otherwise, "no". The interface can activate the cells to fire stored signals along the interlocked sequence, or activate them to fire randomly regardless of existing sequence. The episodic storage receives both symbol and representation inputs from the semantic system as shown in Figure 1, which coincides with the fact that the hippocampus mainly receives inputs from the neocortex (e.g., Aggleton & Brown, 1999; Gluck et al. 2003).

Table 2

In Figure 1, the Mode Selector selects input for the semantic system. The input can be switched between external inputs coming from the *Sin* and *Rin*, and internal inputs coming from the Storage. Furthermore, two other properties of the system are noted. First, the cognitive system in Figure 1 consists of two memory systems: a slow learning semantic system and a fast learning Episodic Storage. The same configuration has been implemented in all the studies, mentioned previously, in the simulations of memory consolidation and retrograded amnesia. Second, episodic memory is initially stored within the Storage before memory consolidation. McClelland et al. (1995) and O'Reilly & Rudy (2000) have implemented a similar approach in their studies.

## 3. Simulations

### 3.1. Episodic learning

Table 3 lists a group of input pairs that are to be learned. Each input pair contains a symbol and its representation. For example, the symbol of the first pair is "III", whose representation has three peaks; the symbol for the fourth pair is "z" which stands for "zero", whose representation has no peak. In other words, the symbol input is the sign for how many peaks exist in its representation. During episodic learning, the input pairs are presented to the system at both *Sin* and *Rin*, one pair at a time with the sequence indicated in the table.

Table 3

Table 4

The semantic system is a slow learner for two reasons. One reason is due to the logically intercorrelated structure. Under this structure, a single memory learns only when it receives a "yes" interlock signal from the single memory next to it. A "yes" interlock signal is one of the necessary conditions for a single memory to learn, as indicated in Table 1. This intercorrelated structure ensures that logically associated concepts are systematically learned based on their original associations. All of the interlock signals have been marked in Figure 1 to show the associations. The other reason is the "modification delay" that delays a single memory to fire a "yes" interlock signal. For example, within a memory triangle, the single memory marked with "2" cannot learn unless the one marked with "1" has already learned and fired a "yes" interlock signal. Because of the slow learning mechanism, all input pairs of Table 3 are unlearned, except for the 6$^{th}$ pair that has the least complicity and is at the bottom of a logical hierarchy. These unlearned pairs are forwarded to the Episodic Storage from either subsystem, and each unlearned input pair is stored in one of the memory cells. Importantly, these pairs are stored in their original sequence of arriving because of the interlock mechanism.

The stored information in the Storage can be recalled and recognized, and Table 4 shows some results of the simulations. The first simulation is an "experience recall" that sequentially recalls a series of past experiences. This recall is activated by the Storage's interface, which sends an "*Ifire*" to the enclosed memory cells one at a time, along the interlocked sequence. As a result, each activated signal is collected by the interface and forwarded to *Sout* and *Rout* via their associated subsystems. This experience recall demonstrates the ability of sequential learning of the Episodic Storage. During a recognition process, an "item" is provided at the *Rin* and is forwarded to the Storage via the representation subsystem to be compared with information stored in all memory cells. When a match is found within a memory cell, the cell fires a "yes" signal of *Orec* (see Table 2) that is then directly forwarded to *Sout*. However, when no match is found, all cells fire "no" of *Orec*. Then the interface sends the "no" to *Sout*. Four simulations of recognition tests are listed in the table.

*3.2. Semantic learning from randomly arriving signals*

After the episodic learning, the semantic system is set to receive randomly activated internal signals and to learn conceptual knowledge from them. During this process, the memory cells, one at a time, is randomly activated to fire its stored input pair by the interface of the Episodic Storage. Since semantic system is a slow learner, semantic learning requires repeated stimulations of information in order to acquire conceptual knowledge (i.e., common features or meanings). In other words, the stored "episodic events" have to be repeatedly activated randomly from the Storage and experienced by the semantic system for a great number of times.

The representation signal of the activated pair is transported to the interface of the representation subsystem and is disassembled into a series of data points by the interface. These data points are further delivered to the associated memory triangles and to single memories to stimulate the learning potential of one common feature. In the meantime, the symbol signal of the pair is transported to the symbol subsystem for symbolic learning.

Which single memory is active in learning, at a given moment, is decided by three factors: the interlocked hierarchal structure, the already learned knowledge, and the common feature that is carried by an incoming signal. There are three common features that are carried by those input pairs. They are: (1) whenever there is no peak, a representation is always named "Z", regardless of the length of a line; (2) whenever one peak occurs, the representation is always named "I", regardless of the width of a pit; (3) when the number of peaks in a representation increases, the symbol for representation increases accordingly. The semantic system is constructed to be able to abstract and learn the meanings of line, peak and peaks, while omitting differences in the length of line or width of peak. Among these three common features, the feature about "zero" or "nothing" is the background knowledge of other features and has to be firstly learned. The inputs about "one" carry two common features: one is "zero" and the other is "one". Therefore, they can only be learned after the system is able to utilize the background knowledge of "nothing" to abstract the common feature of "one". The remaining inputs carry three common features of "nothing", "one" and "tally", thus they can only be learned after "nothing" and "one" has been learned. After the three common features have been learned, the background feature of "zero" is stored in *MTr1*, the feature of "one" is stored in *MTr2*, and the feature of "tally" is stored in *MTr3*. In the meanwhile, similar learning process has also occurred in the symbol subsystem: the symbol "Z" is learned by *MTs1*, and the symbols of "I', "II" and "III" are learned by *MTs2*.

Figure 2

After the semantic system has learned all the conceptual knowledge that is carried by those input pairs, it is able to generate any one of the symbols in Table 3, when it is prompted by the associated representation input in the absence of the Storage. Thus, it can be said that all the experienced events during episodic learning have been consolidated into the semantic system. Figure 2 shows that how the semantic learning gradually progresses. Each data point in Figure 2 indicates how many symbol inputs can be regenerated by the semantic system after a given number of random firings. Since the firings are activated at a fixed time interval, Figure 2 also represents a relationship between consolidation rate and time. Similar correlation between consolidation rate and time has been simulated to explain retrograded amnesia due to hippocampal lesions (e.g., Alvarez & Squire, 1994; McClelland *et al*., 1995), because episodic memory for past experience happened earlier should have more chance to be consolidated than the one happened later.

### *3.3. Preserved recognition and semantic knowledge, and impaired experience recall*

Severely impaired delayed recall, e.g., picture drawing (Vargha-Khadem et al. 1997) and storytelling (Isaacs et al. 2003), preserved recognition, and preserved semantic learning capacities are the three characteristics of developmental amnesia.

In this study, the impairment of delayed recall is proposed to be the result of impaired sequential/spatial learning. In the presented model, the sequential learning capacity is realized by the interlock mechanism and its recall is executed by the interface of the Episodic Storage. After the interlock mechanism is impaired, only small portion of stored events is registered in the interface correctly, and other events are registered incorrectly. In other words, the association among the stored events is almost lost, and Storage is only able to activate those correctly registered events, but not others. As a result, When performing "experience recall" task, only small portions of experienced events are activated and recalled, and the recalled components are alternated by blanks because of the failed attempts to activate nonexisting memory cells. In here, the experience recall is thought to be equivalent to picture drawing or storytelling, because both involve the recall of an entire experience of sequenced events or associated items. A simulation of impaired experience recall is shown in Table 5, which is part of the successful experience recall listed in Table 4.

Table 5

The system with impaired sequential learning of the Episodic Storage is tested for recognition tasks, and the recognition simulations are also listed in the table. During recognition test, an experienced event (representation input) is provided to the *Rin*. Similar to episodic learning process, the input is forwarded to the interface of the Storage via the representation subsystem. The interface delivers the input to its enclosed memory cells in parallel. Each of the cells compares the incoming signal with what has been stored. When a match is found, the cell fires a "yes" signal of *Orec* (see Table 2), otherwise "no", and this signal is collected by the interface and forwarded directly to the *Sout*. Both experience recall and recognition simulations are performed before memory consolidation.

With the impaired sequential learning ability of the Storage, the system is set to go through memory consolidation process after episodic learning. After the consolidation is finalized, the system is tested in terms of "object naming" in absence of the Episodic Storage. During the test, an object of representation signal is presented at the *Rin*, which is forwarded to the interface of the representation subsystem. The interface disassembles the input into a series of data points and delivers them to the enclosed single memories for comparison of common features. Whenever a match is found, the associated single memory fires an excitation signal (*Oexc*, see Table 1). All activated excitation signals are collected and organized by the interface into a chunk and forwarded to the symbol subsystem via the "bundle of internal signals". The chunk is then disassembled by the second interface into a series of excitation signals, which are delivered to all single memories to excite associated symbols stored (see also Table 1). In the end, all excited symbols are collected and forwarded to *Sout* as a symbol output.

The three simulations in Table 5 indicate that the system is able to name a presented item correctly after the memory consolidation. Furthermore, the semantic system is able to tally peaks in an unfamiliar representation input that has not been experienced during episodic learning. Such capacity is often named as "flexibility" because the acquired conceptual knowledge is resistant to the alternations of cues, which is one of the two key characteristics of semantic knowledge (e.g., O'Kane, et al., 2004). The reason that the impaired sequential learning has no influence to the outcome of memory consolidation is because memory consolidation is featured with random activation and no sequential access of information is necessary.

**4. Discussion**

*4.1. Memory consolidation*

The presented model consists of a slow learner (the semantic system) and a fast learner (the Episodic Storage). Past experience is stored first in the Storage after a single exposure. Then, the semantic system acquires conceptual knowledge slowly from randomly activated past experience stored in the Storage. After the acquisition process, the Storage is removed. In the absence of the Storage, the semantic system is still able to respond correctly to any already experienced "event". This coincides with a popular view that memory consolidation is the process by which memory becomes independent of the MTL and only dependent of the neocortex. Thus the process can be seen as the memory consolidation process of the cognitive system. This acquisition process has a number of characteristics. They are: (1) episodic memory is initially stored in the Storage, (2) the stored experience is randomly activated, (3) the acquisition may only be realized after numerously repeated encounters of the activated information, and (4) the acquired knowledge is conceptual knowledge that is one of the properties of semantic memory.

The four characteristics have either been revealed in empirical studies, or proposed in theoretical studies and implemented in some existing computational models. Halgren (1984) and Teyler & DiScenna (1986) suggested that episodic memory might initially be stored in the hippocampus, and McClelland et al. (1995) and O'Reilly & Rudy (2000) implemented the initial episodic storage in their computational modeling of memory consolidation and retrograde amnesia. However, Squire et. al. (1984) and Murre (1996) viewed that episodic memory might be stored in the neocortex initially and the MTL served as the link to bond the stored episodic segments.

Memory consolidation is viewed as a gradual process in which episodic memory is gradually incorporated into an existing framework in the neocortex, or slowly teaches the hippocampal representations into the neocortex. Being a gradual process, memory consolidation has to occur constantly, and memories stored in the hippocampus have to be continually revived. Squire and Alvarez (1995) have argued that if memory consolidation is occurring constantly, dreaming may be the most suitable process to explain why the process of consolidation does not regularly intrude into our consciousness. Furthermore, memory consolidation has also been simulated with random activation of cortical connections via the hippocampus (Alvarez & Squire, 1994; Murre, 1996), or random activation of stored memories in the hippocampus (McClelland, et al., 1995, Zhang, 2009). The same role of the hippocampus in memory consolidation is also concluded from the findings of hippocampal replaying of recent waking patterns in rats (e.g., Pavlides & Winson, 1989) and in humans (Staba, et al., 2002) during sleep. Interestingly, random activation has been considered by many researchers to be a key characteristic of dreaming since the initial study reported by Hobson & McCarley (1977). Due to the nature of parallel distributed processing of connectionist modeling, one neuron in one subsystem may have to connect with many neurons in a different subsystem, thus a single stream of information flow that is compatible with dreams would not exist in the simulated processes of memory consolidation. As a comparison, there are two equivalent information pathways from the Episodic Storage to the semantic system in this study. Randomly activated information flow can be entirely recorded that is compatible with "dream report", and has been shown in the previous study of dream learning (Zhang, 2009). The results of the memory consolidation are also demonstrated in this study, and the progress of the consolidation process is given in Figure 2.

The consolidation curve in Figure 2 agrees with the view that memory consolidation is a slow and gradual process of learning, and shows a positive correlation between consolidated items and consolidation period or the number of encounters. This curve has the same trend as Figure 12b of a consolidation simulation by McClelland, et al. (1995), and looks very much like the correlation between "maximum performance" and "consolidation period" found in rats whose hippocampuses were purposely lesioned in certain given days after they were trained (Kim & Fanselow, 1992).

Although semantic memory is defined as generic and factual knowledge that is featured with flexibility and awareness (Tulving, 1972), and is considered to be associated with memory consolidation (see Eichenbaum, 2004), the characteristic of flexibility has not been demonstrated in existing consolidation simulations. The cued recall (McClelland, et al., 1995) of a word-string only demonstrates that word-association can be consolidated from a faster learner to a slow learner that represents the neocortex. The cued recall of patterns (Alvarez & Squire, 1994; Murre, 1996) only demonstrates that the association between patterns can be independent of the MTL after consolidation. On the other hand, the simulations given in Table 5 show that the acquired knowledge is flexible, and the semantic system can utilize its acquired knowledge to process unfamiliar stimuli and respond to them correctly.

The semantic system of the computational model consists of a pair of subsystems, and each of which has its own learning process and chain of memories. This construct coincides with the idea of split-brain that suggests each brain half appears "to have its own, largely separate, cognitive domain", and to "have its own learning processes and its own separate chain of memories". Sperry in his Nobel lecture (1982) also noted that our left hemisphere is capable of comprehending printed and spoken word, and our right hemisphere is word-deaf and word-blind, but capable of comprehending spatial and imagistic information. The idea of split-brain is also supported by many recent empirical studies. For example, in most cases of semantic dementia and anomia, the atrophy is typically marked on the left hemisphere, or with left predominance (e.g., Hodges et al., 1995; Mummery, 2000; Ralph, et al., 2001). On the other hand, patients with prosopagnosia (face blindness, Evans, et al., 1995), and semantic dementia who have more deficit in comprehending meanings (Ralph, et al., 2001) are accompanied with right predominant atrophy. Similar association between impairment pattern and hemispherical asymmetry of lesion can also be found in many priming studies of patients with brain damages.

*4.2. Developmental amnesia*

Developmental amnesia is an atypical form of memory deficit that has been discovered to occur in children with hippocampal damage. A clear dissociation has been revealed between relatively preserved semantic memory and badly impaired episodic memory with damage that is restricted mainly to the hippocampus. A group of characteristics of this impairment is summarized as following. Patients always suffer bilateral damage to the hippocampal formation at very early ages with sparing of surrounding cortical areas (Vargha-Khadem et al. 1997). Such impairment becomes evident when the hippocampal atrophy is as low as 27 percent compared with normal subjects, and the highest atrophy found is about 56 percent (Isaacs et al. 2003). Such patients have equivalent IQ, digit span and Corsi block span as that of control groups, but are impaired relative to control groups on nearly all delayed memory measures on a wide variety of verbal and visual tasks. They score anywhere from a few percent up to about 25% of the control groups in both delayed story recall and delayed reproduction of geometric designs. However, their recognition ability appears to be normal or close to normal (Holdstock et al., 2002; Vargha-Khadem et al. 1997).

In short, developmental amnesia is characterized with (1) preserved recognition ability, (2) equivalent semantic knowledge as that of controls presumably due to preserved semantic learning capacity, and (3) badly but not entirely impaired recall ability when performing delayed storytelling and picture reproduction. These three characteristics are accompanied by the 27-56% hippocampal atrophy.

It seems that such early loss of episodic memory may impede cognitive development and result in severe mental retardation, since many have believed that semantic memory is mainly acquired from episodic memory through memory consolidation. In order to interpret the seemingly conflict between impaired episodic memory and spared semantic learning, different explanations have been offered. Vargha-Khadem et al. (1997), Baddeley et al. (2001) and Isaacs et al. (2003) agreed that the recollective process of episodic memory is not necessary either for recognition or for the acquisition of semantic knowledge. This explanation is based on the view that recall and recognition can be dissociated, and on the view that episodic memory is partially dissociated from semantic memory, but is fully dependent on the hippocampus. Squire and Zola (1998), on the other hand, suggested a number of different explanations. One explanation is that patients with a damaged hippocampus at a young age may develop some form of adaptation to learn semantic knowledge. This idea is discounted by a carefully designed study about the ages of the amnesic onset (Vargha-Khadem et al. 2003). Another explanation is that since none of the patients have entirely lost their "recall memory", the residual "recall memory" may be enough to explain the near normal semantic memory performance, although no detailed mechanism was offered.

In Table 5 of this study, a similar set of three characteristics has been simulated and shown, including badly but not entirely impaired experience recall and intact recognition and semantic learning. These three characteristics result from the damaged sequential or spatial learning, while other functions of the Episodic Storage are still preserved. In the impaired experience recall, only a small portion of the past experience is recalled (compared with the successful recall in Table 4). This partial recall is comparable to the data given in the initial study of Vargha-Khadem et al. (1997). The recalled materials can be as high as about 25% (patient Jon) in delayed story recall and about 20% (patient Beth) in delayed reproduction of geometric designs, in comparison to the performance of controls. The impaired experience recall is very much similar to the reprints of the delayed recalls of a geometric design performed by three patients in the study of Vargha-Khadem et al. (1997). The geometric design was a single structure consisting of many interlaced triangles, rectangles, and lines. The patients were only able to redraw a small portion of the whole design. Interestingly, the redrawn portions were mostly detached triangles and rectangles and the associations among the patterns were lost. The same feature of detached "items" is also shown in the impaired experience recall.

The damaged sequential learning mechanism does not necessarily impair the recognition ability of the Storage, because recognition utilizes the comparison function that is a different mechanism from sequential learning. Thus, information can be recognized as long as it has been stored. The three simulations show that experienced events can always be recognized (the two recognition tests that generated "yes" output in Table 5), while inexperienced events can not be (the one recognition test that generated "no" output). Furthermore, semantic learning is thought to be associated with random activation of stored experience through memory consolidation process, not with sequential retrieval of stored information. Thus, damaged sequential learning function should not affect semantic learning, which has also been simulated and shown in Table 5.

The occurrence of these three characteristics is reproduced in simulations based on the assumption that only sequential learning ability is lesioned while the storage function of the hippocampus is not affected in developmental amnesia. The impaired recall of sequenced experiences is resulted from the impaired associations among these experiences when they are stored. Two mechanisms can cause such impaired association: either the Storage has basically lost the capacity to encode the associations, or the associations have been encoded wrongly.

This combination of one damaged function and spared other functions coincides with the fact that developmental amnesia is only found in patients, so far, who had hippocampal atrophy ranged from 27 to 56 percent. On the other hand, Isaacs et al. (2003) also reported a preterm group who had no obvious deficits in both episodic memory and semantic memory and whose hippocampal atrophy could be as high as 23%. It seems unreasonable if developmental amnesic patients had lost their entire function of hippocampus, when the atrophy had increased for a small percentage compared with the preterm group.

## 5. Conclusion

In summary, this study presents a cognitive and computational model that consists of an Episodic Storage and a semantic system. The Storage is proposed to resemble the hippocampus for the functions of storage and sequential learning. Semantic learning from episodic memory through memory consolidation is simulated. The acquired knowledge is demonstrated for the first time to be flexible and resistant to the alternations of cues. The unique characteristics of developmental amnesia are simulated and reproduced after the sequential learning capacity is damaged. As a result, sequential retrieval of stored experience is severely impaired, while both recognition ability and semantic learning are spared. This study is in agreement with one of the explanations offered by Squire and Zola (1998) in that the "residual" ability of the hippocampus in developmental amnesia may be enough to support semantic learning. The simulations also agree with the view that recognition memory can be dissociated from recall memory.

26Zola, S.M., Squire, L.R., & Amaral, D. (1986). Human amnesia and the medial temporal region: enduring memory impairment following a bilateral lesion limited to field CA1 of the hippocampus. *J Neurosci., 6,* 2950-2967.26

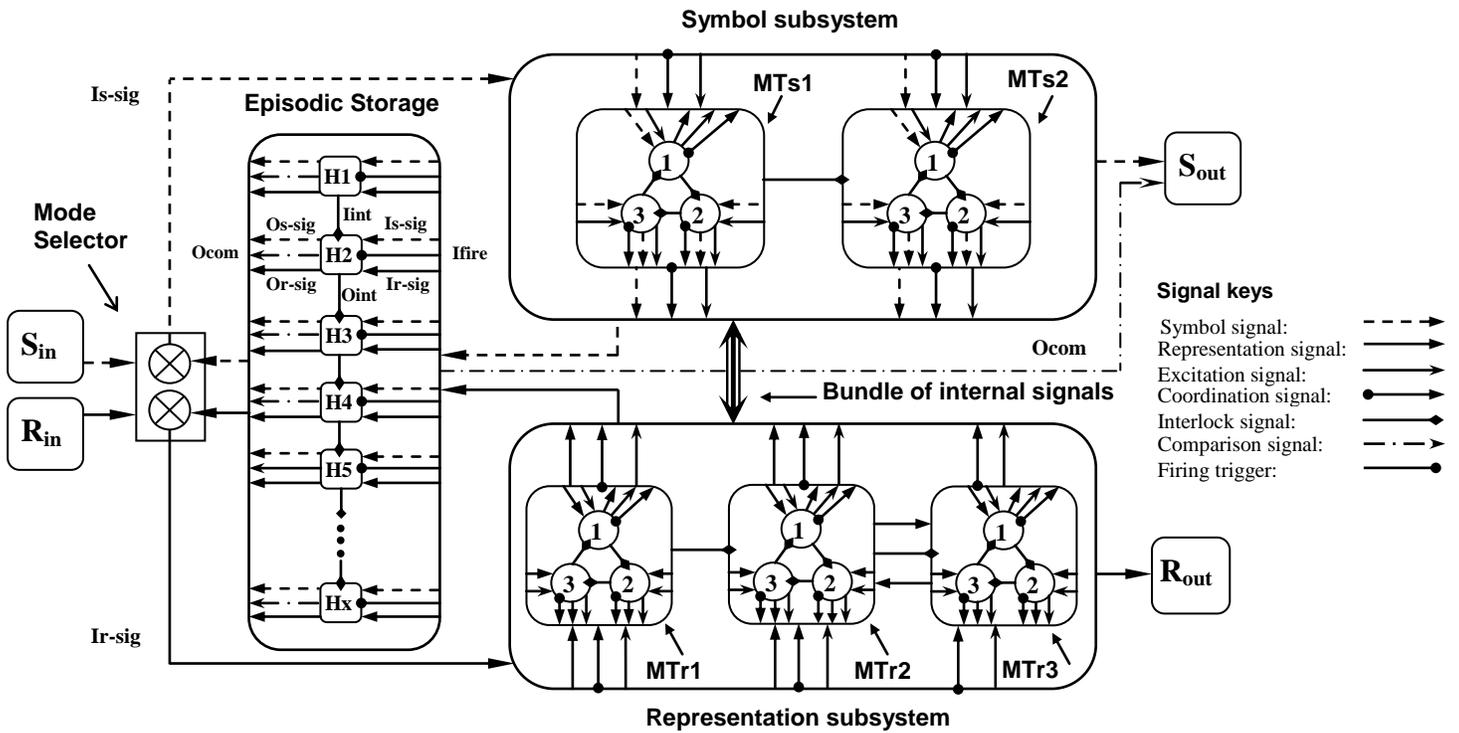

**Figure 1**. The cognitive structure of a computational system that consists of a semantic system (the combination of the two subsystems) and an Episodic Storage. Each of the subsystem has three cognitive layers—single memory (each of the small circles marked with 1, 2 and 3) whose function states are given in Table 1, memory triangle (*MTxx*, consisting of three single memories), and subsystem (consisting of two or three memory triangles). The symbol subsystem acquires symbols, and the representation subsystem acquires common features or meanings. Conceptual knowledge is learned when an acquired symbol is associated with an acquired common feature. The Episodic Storage consists of a number of memory cells whose function states are given in Table 2. All of the cells are interlocked by interlock signals, and the Storage can store and retrieve a sequence of experiences. The Storage can also retrieve stored information randomly without utilizing the sequential learning ability.

During episodic learning, a signal input (*Is-sig*) and a representation input (*Ir-sig*) are presented at *Sin* and *Rin*, respectively. Since the semantic system cannot learn the inputs immediately, it forwards them to the Storage for immediate storage. During experience recall, the Storage activates a series of stored information along the interlocked sequence. The activated series of either *Os-sig* or *Or-sig* is forwarded to the *Sout* via the symbol subsystem, or the *Rout* via the representation subsystem. During recognition process, a representation input is presented at *Rin* and is forwarded to the Storage for comparison, again via the representation subsystem. A "*yes*" or "*no*" signal of *Ocom*, as the result of comparison, is projected to the *Sout*. During memory consolidation, the stored information is randomly and repeatedly activated from the Storage, and is forwarded to the semantic system via the Mode Selector, to stimulate semantic learning.

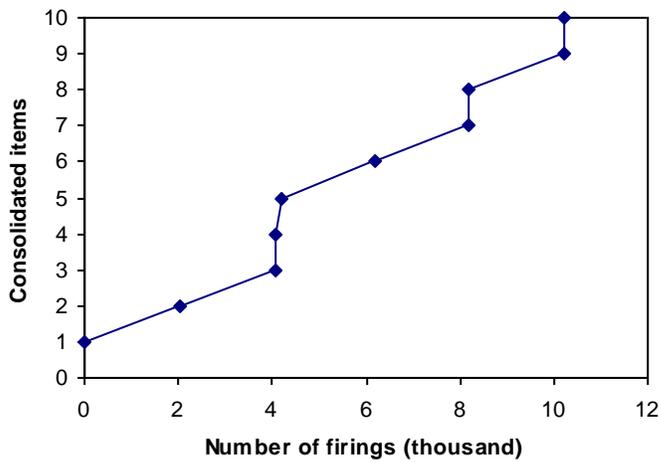

**Figure 2**. A simulation of memory consolidation showing the relationship between consolidated items and the number of random firings. Here, the ratio of T/t (modification delay/random firing interval) is set at 2000.

**Table 1.** Three important states of a single memory, after [29]

| State | | Input | Output |
|---|---|---|---|
| Learning | Step 1: firing Ocor signal | Isig = "Io"<br>Iexc = null<br>Iint = "yes" | Osig = null<br>Oexc = null<br>Oint = "no"<br>Ocor = "yes" |
| | Step 2: storing "Io" and "Iexco" permanently | Isig = "Io"<br>Iexc="Iexco"<br>Iint = "yes" | Osig = null<br>Oexc = null<br>Oint = "yes"<br>Ocor = "yes" |
| Firing stored "Io" upon receiving "Iexco" after the single memory has learned. | | Isig = any<br>Iexc="Iexco"<br>Iint = "yes" | Osig = "Io"<br>Oexc ="Iexco" or null*<br>Oint = "yes"<br>Ocor = "no" |
| Firing stored "Iexco" upon receiving "Io" after the single memory has learned. | | Isig = "Io"<br>Iexc = any<br>Iint = "yes" | Osig = "Io" or null**<br>Oexc = "Iexco"<br>Oint = "yes"<br>Ocor = "no" |

\* Depending on Isig
\*\* Depending on Iexc

**Table 2.** Three function states of a memory cell

| State | Input | Output |
|---|---|---|
| Learning:<br>To store "Iso" and "Iro" | Is-sig = "Iso"<br>Ir-sig = "Iro"<br>Iint = "yes"<br>Ifire= "no" | Os-sig = null<br>Or-sig = null<br>Oint = "yes"<br>Oreco = null |
| Firing:<br>To fire stored "Iso" and "Iro" | Is-sig = null<br>Ir-sig = null<br>Iint = null<br>Ifire = "yes" | Os-sig = "Iso"<br>Or-sig = "Iro"<br>Oint = null<br>Oreco = null |
| Comparing:<br>To compare incoming signal with stored "Iso" and "Iro" | When Is-sig = "Iso"; other signals = null | Oreco = "yes"; Or-sig = "Iro"; other signals = null |
| | When Ir-sig = "Iro"; other signals = null | Oreco = "yes"; Os-sig = "Iso"; other signals = null |
| | When Is-sig ≠ "Iso"; or Ir-sig ≠ "Iro"; other signals = null | Oreco = "no"; other signals = null |

**Table 3.** Input pairs for episodic learning

|   | Symbol | Representation |
|---|---|---|
| 1st | III | |
| 2nd | I | |
| 3rd | I | |
| 4th | Z | |
| 5th | IIII | |
| 6th | Z | |
| 7th | Z | |
| 8th | I | |
| 9th | II | |
| 10th | Z | |

**Table 4.** Simulations of experience recall and recognition after episodic learning

| Task | Input | Output |
|---|---|---|
| Experience recall | | III I I Z IIII Z I II Z |
| Recognition | | yes |
| | | yes |
| | | yes |
| | | no |

**Table 5.** Simulations of impaired experience recall, and intact recognition and semantic learning

| Condition | Input | Output | Comment |
|---|---|---|---|
| Tested after episodic learning, but before consolidation | Experience recall mode | I    Z | Impaired experience recall |
| | | yes | Intact recognition |
| | | yes | Intact recognition |
| | | no | Intact recognition |
| Tested after consolidation and in absence of the Storage. | | Z | Intact tally |
| | | II | Intact tally |
| | | IIIII | Intact tally |